# High-Entropy Hydrides for Fast and Reversible Hydrogen Storage at Room Temperature: Binding-Energy Engineering via First-Principles Calculations and Experiments

Abbas Mohammadi[1], Yuji Ikeda[2], Parisa Edalati[1], Masaki Mito[3], Blazej Grabowski[2], Hai-Wen Li[4] and Kaveh Edalati[1,*]

[1] WPI, International Institute for Carbon-Neutral Energy Research (WPI-I2CNER), Kyushu University, Fukuoka 819-0395, Japan
[2] Institute for Materials Science, University of Stuttgart, Pfaffenwaldring 55, 70569 Stuttgart, Germany
[3] Graduate School of Engineering, Kyushu Institute of Technology, Kitakyushu 804-8550, Japan
[4] Hefei General Machinery Research Institute, Hefei 230031, China

Despite high interest in compact and safe storage of hydrogen in the solid-state hydride form, the design of alloys that can reversibly and quickly store hydrogen at room temperature under pressures close to atmospheric pressure is a long-lasting challenge. In this study, first-principles calculations are combined with experiments to develop high-entropy alloys (HEAs) for room-temperature hydrogen storage. $Ti_xZr_{2-x}CrMnFeNi$ ($x$ = 0.4-1.6) alloys with the Laves phase structure and low hydrogen binding energies of -0.1 to -0.15 eV are designed and synthesized. The HEAs reversibly store hydrogen in the form of Laves phase hydrides at room temperature, while (de)hydrogenation pressure systematically reduces with increasing the zirconium fraction in good agreement with the binding energy calculations. The kinetics of hydrogenation are fast, the hydrogenation occurs without any activation or catalytic treatment, the hydrogen storage performance remains stable for at least 1000 cycles, and the storage capacity is higher than that for commercial $LaNi_5$. The current findings demonstrate that a combination of theoretical calculations and experiments is a promising pathway to design new high-entropy hydrides with high performance for hydrogen storage.

* Corresponding author: K. Edalati (E-mail: kaveh.edalati@kyudai.jp; Tel: +81-92-802-6744)

# 1. Introduction

Excessive consumption of fossil fuels and $CO_2$ emission caused by their utilization have led to the crisis of global warming. Nowadays, finding clean fuels which do not emit $CO_2$ is a serious challenge for scientists and industry leaders. Hydrogen is the cleanest fuel and has attracted attention as a substitute for fossil fuels; however, besides the necessity for clean production of hydrogen, its safe and compact storage is a significant challenge [1]. Storage of hydrogen in the form of gas and liquid is conventionally used for various applications. However, the amount of stored hydrogen in the form of gas in typical commercial tanks with 225 liters and 20 MPa pressure is just 4 Kg [2]. Therefore, this typical method has limitations in terms of volumetric and gravimetric storage densities, although there are recent trends to increase the storage pressure to 70 MPa by using special tanks [1]. In the liquid storage method, the volumetric and gravimetric storage densities are higher, and the safety is better than for the gas storage method, but liquifying hydrogen at low temperatures makes the method expensive and evaporation losses can also occur [2]. Solid-sate hydrogen storage particularly in the form of metal hydrides provides the most compact and safest technology to store hydrogen [2].

To realize the application of metal hydrides for hydrogen storage, they should have several features such as the capability for reversible absorption and desorption of hydrogen at ambient temperature, high cycling stability, fast kinetics, appropriate storage pressure near atmospheric pressure and high gravimetric capacity (high gravimetric capacity for stationary applications is not as critical as it is for mobile applications) [3]. Magnesium hydride and complex hydrides are well-known materials with high storage capacity, but they suffer from high thermodynamic stability, and thus, they function only at high temperatures [1]. So far, a limited number of materials such as TiFe ($TiFeH_2$) and $LaNi_5$ ($LaNi_5H_6$) have been introduced for room-temperature hydrogen storage [3-6], but they exhibit other shortcomings such as the activation problem in TiFe, and a high price and low storage capacity in $LaNi_5$ [3-6]. Therefore, there are still significant demands to introduce new metal hydride systems that can satisfy the requirements for hydrogen storage at ambient temperature.

High-entropy materials, which are solid solutions of at least five principal elements with a configurational entropy higher than $1.5R$ ($R$: gas constant), have attracted attention in recent years for various applications including hydrogen storage [7]. The presence of several elements in a single phase allows to manipulate the electronic structure, hydrogen binding energy and accordingly hydrogen storage temperature and pressure by careful selection of principal elements and their concentrations [7]. TiVZrNbHf [8-10], TiVCrNbMo [8], TiVCrNbTa [8], $Ti_{0.2}Zr_{0.2}Hf_{0.2}Mo_{0.1}Nb_{0.3}$ [11], $Ti_{0.2}Zr_{0.2}Hf_{0.2}Mo_{0.2}Nb_{0.2}$ [11], $Ti_{0.2}Zr_{0.2}Hf_{0.2}Mo_{0.3}Nb_{0.1}$ [11], TiZrVNbCr [12], $V_{30}Ti_{30}Cr_{25}Fe_{10}Nb_5$ [13], $V_{35}Ti_{30}Cr_{25}Fe_5Mn_5$ [13], $Mg_{0.10}Ti_{0.30}V_{0.25}Zr_{0.10}Nb_{0.25}$ [14], TiZrNbFeNi [15], TiZrNbCrFe [16], MgAlTiFeNi [17], $Al_{0.10}Ti_{0.30}V_{0.25}Zr_{0.10}Nb_{0.25}$ [18], $Mg_{12}Al_{11}Ti_{33}Mn_{11}Nb_{33}$ [19], MgVAlCrNi [20], MgVTiCrFe [21], AlCrFeMnNiW [22], TiZrHfScMo [23], $MgZrTiFe_{0.5}Co_{0.5}Ni_{0.5}$ [24] and LaNiFeVMn [25] are some of the HEAs which have been investigated for hydrogen storage. However, as discussed in a recent review paper [7], these HEAs have drawbacks such as either high-temperature requirement for hydrogen storage, poor hydrogen storage reversibility, poor activation, or high storage pressure [8-25], which limit their applications. Although the research on high-entropy hydrogen storage materials is still in its early stages, designing these alloys by theoretical and computational methods is expected to provide a pathway to discover new materials that can quickly and reversibly store hydrogen under ambient conditions.

In this work, first-principles calculations are combined with experiments to design HEAs for room-temperature hydrogen storage. The designated alloys, $Ti_xZr_{2-x}CrMnFeNi$ ($x = 0.4-1.6$) with the Laves phase structure and low hydrogen binding energies of -0.1 to -0.15 eV, show fast and reversible hydrogen storage at ambient temperature under pressures adjustable to the atmospheric pressure by changing the amounts of titanium and zirconium. This simultaneous application of theoretical and experimental studies to high-entropy hydrogen storage materials confirms the significance of this strategy in developing new HEAs that can satisfy the requirements for stationary hydrogen storage applications.

## 2. Materials and methods
### 2.1. Empirical material design

The key issue in designing room-temperature hydrogen storage materials is to adjust the hydrogen binding energy to a negative value close to zero [26]. An earlier study on first-principles calculations of Mg-based alloys suggested that binding energies of about -0.1 eV per hydrogen atom can be an appropriate target to achieve room temperature hydrogen storage [26]. Moreover, the hydrogen binding energy should be slightly more negative than -0.1 eV to reduce the equilibrium hydrogen storage pressure close to ambient pressure. Although such a concept has not been used to design HEAs so far, three empirical criteria were suggested by the current authors to achieve hydrogen storage at low temperatures in HEAs [27]: (i) $AB_2$-type atomic configuration (A: elements which react with hydrogen; B: elements with low affinity with hydrogen), (ii) C14 Laves phase structure formation in the alloy and hydride; and (iii) valence electron concentration (VEC) of 6.4.

- Hydrogen storage materials are usually a mixture of A-type elements (such as lanthanum, magnesium, titanium, etc.) and B-type elements (such as nickel, iron, manganese, etc.) [1]. The A-type elements have negative hydrogen binding energies and produce stable hydrides, while B-type elements have positive binding energies and usually do not absorb hydrogen, as schematically shown in Fig. 1a using the data reported in the literature [3,26,28]. Therefore, a combination of A-type and B-type elements can lead to the formation of alloys with an appropriately low hydrogen binding energy for room-temperature hydrogen storage such as TiFe and $LaNi_5$ [1]. In this regard, it was found that $AB_2$-type HEAs have a high potential for low-temperature hydrogen storage [27], while AB-type and $A_3B_2$-type systems are other candidates, yet with less potential [15,16].
- The Laves phase alloys are considered as potential materials for hydrogen storage with high cycling stability for reversible hydrogenation and dehydrogenation and fast kinetics [29]. Another benefit of Laves phase alloys is that they can have lower cost compared to rare-earth-based alloys such as $LaNi_5$ [29]. It was shown by both experiments and the CALPHAD (calculation of phase diagrams) method that the Laves phases can be formed in some high-entropy hydrogen storage systems such as Ti-Zr-Cr-Mn-Fe-Ni [27], Ti-Zr-Nb-Fe-Ni [15] and Ti-Zr-Nb-Cr-Fe [16].
- VEC is another key parameter that can be considered in designing alloys for reversible hydrogen storage at room temperature. Metals with low VEC such as lithium, magnesium and titanium usually produce stable hydrides, which can release hydrogen only at high temperatures, while metals with high VEC such as cobalt, nickel and copper usually exhibit low affinity with hydrogen. It was suggested that HEAs with a VEC value of 6.4 can desorb hydrogen at temperatures close to room temperature [8]. Although adjusting VEC is hard in simple binary or ternary alloys, it can be adjusted much easier in HEAs by changing the type and fraction of the principal elements.
- The hydrogen binding energy is the most important parameter that needs to be adjusted for reversible hydrogen storage at room temperature and under atmospheric pressures. The dehydrogenation temperature increases and the equilibrium hydrogen pressure decreases with increasing the absolute value of negative binding energy. Only a limited number of materials such as LaNi$_5$ and TiFe show appropriate hydrogen binding energies for room-temperature hydrogen storage. As schematically shown in Fig. 1a, the target is to set the hydrogen binding energy to a value slightly more negative than -0.1 eV by adjusting the fraction of elements [26].

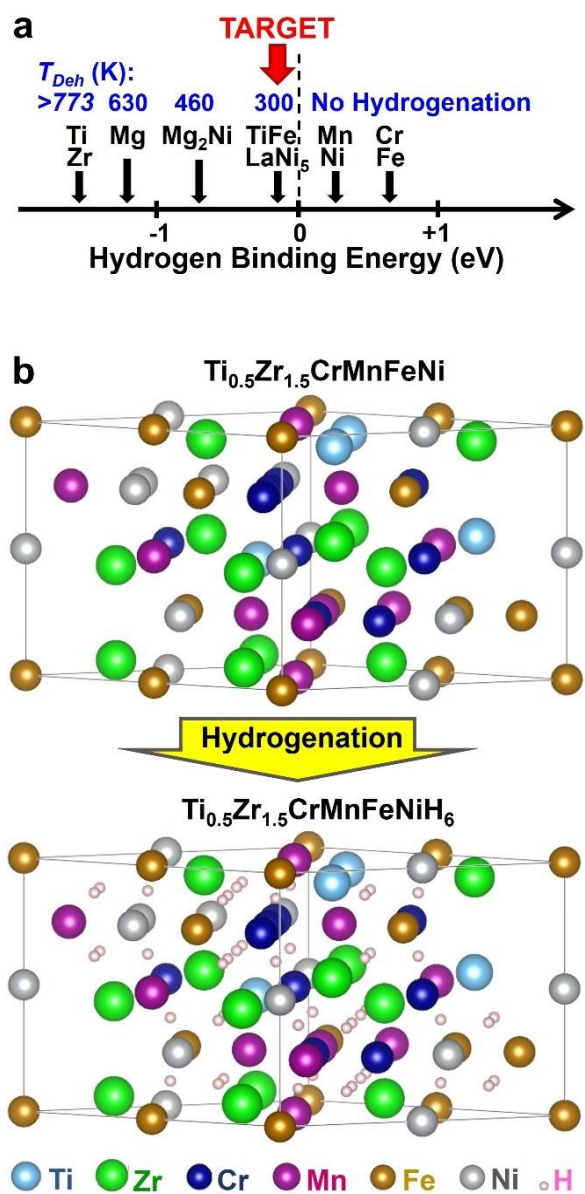

Fig. 1. (a) Schematic illustration of hydrogen binding energy on hydrogenation and dehydrogenation of different materials. (b) One of SQS models of HEA Ti$_{0.5}$Zr$_{1.5}$CrMnFeNi and corresponding hydride Ti$_{0.5}$Zr$_{1.5}$CrMnFeNiH$_6$ (visualized using VESTA [33]), considered in present *ab initio* simulations.

Therefore, this study focuses on HEAs with the AB$_2$-type Laves phase structure and a VEC value of 6.4; and among the available options, the Ti-Zr-Cr-Mn-Fe-Ni system can fit all these requirements [27]. The hydrogen binding energy is adjusted by changing the fraction of zirconium in balance with titanium, because zirconium has a larger atomic radius than the other elements in this alloying system, and thus, adjusting its fraction is supposed to have the most significant effect on lattice volume and binding energy. Three compositions of $Ti_xZr_{2-x}CrMnFeNi$ ($x$ = 0.5, 1.0, 1.5) were theoretically studied by first-principles binding energy calculations and four compositions of $Ti_xZr_{2-x}CrMnFeNi$ ($x$ = 0.4, 0.8, 1.2, 1.6) were experimentally examined by hydrogen storage characterizations.

**2.2. First-principles calculation methods**

**2.2.1. Crystal structure modeling of alloys**

The $AB_2$-type hexagonal C14 Laves phases have the space group of $P6_3/mmc$ (No. 194), in which the A-type atoms occupy the 4$f$ Wyckoff sites, and the B-type atoms occupy the 2$a$ and the 6$h$ Wyckoff sites [29,30]. These phases have in total 12 atoms in their unit cell. If A and B atoms in the $AB_2$-type Laves phases are approximated by close-packed rigid spheres with the radii $r_A$ and $r_B$, respectively, the ratio $r_A/r_B$ is given by $(3/2)^{1/2} \approx 1.225$ [30]. These atomic positions in the close-packing case, which are considered the ideal positions, lead to a $c/a$ ratio of $(8/3)^{1/2} \approx 1.633$ ($a$ and $c$ are lattice parameters).

The C14 $Ti_xZr_{2-x}CrMnFeNi$ Laves phase alloys ($x$ = 0.5, 1.0, 1.5) were modeled using 48-atom supercells with a 2 × 2 × 1 expansion of the primitive cell of the C14 phase. These three compositions, which slightly differ from the experimental compositions, were selected to investigate the dependence of the hydrogen binding energy on the titanium fraction with 48-atom systems and reasonable computation time. The A sites were occupied by titanium and zirconium, and the B sites were occupied by chromium, manganese, iron, and nickel. The sublattice chemical disorder was modeled using special quasi-random structure (SQS) configurations [31]. Correlation functions of the first several nearest-neighbor doublet, triplet, and quartet clusters were optimized to be close to the ideal values of fully random configurations using the simulated annealing approach implemented in the ICET code [32]. To achieve better statistics, six different configurations were considered for each composition by permuting the elements. Fig. 1b (upper image) shows one example of modeled structure of a HEA which was modeled using SQS [31] and visualized using VESTA [33].

**2.2.2. Crystal structure modeling of hydrides**

To model the structure of the high-entropy hydrides, crystallographic information and reports from the literature were considered. In the Laves phase alloys, there are 17 tetrahedral interstitial sites per formula unit $AB_2$, which could be occupied by hydrogen atoms [34]. These interstitial sites are surrounded by either four B atoms ($B_4$), one A and three B atoms ($AB_3$), or two A and two B atoms ($A_2B_2$). For many conventional Laves phases composed of the elements as utilized in the present study, previous *ab initio* simulations found that the $A_2B_2$ sites are the most energetically preferable sites for hydrogen atoms [35-39], and indeed many experiments found hydrogen atoms at the $A_2B_2$ sites [40-48]. This may be intuitively understood because the $A_2B_2$ sites have larger volumes than the $B_4$ and $AB_3$ sites when the atoms A and B are on the ideal Laves lattice sites, as summarized in Table 1. The hydrogen atoms in Laves phases should also be repulsive to each other and should not occupy very close interstitial sites, as empirically suggested [49] and confirmed by first-principles calculations for some Laves phases [36,39].

Since previous experimental studies reported that the $AB_2$-type high-entropy Laves phase produces a Laves phase hydride with the composition $AB_2H_3$ [27], the same hydrogen fraction was considered for modeling in the present study. It should be noted that for the present 48-metal-atom supercell models, there are in total 272 interstitial sites, all of which have different local chemical environments. Therefore, even for the fixed composition AB$_2$H$_3$, there are in total over 10$^{53}$ possible ways of hydrogen occupations, which are obviously impossible to test in a brute-force manner. Therefore, based on the previous experiments and simulations mentioned above [34-48], we *a priori* assumed that all hydrogen atoms occupy the A$_2$B$_2$ sites. Moreover, while there are four symmetrically inequivalent A$_2$B$_2$ sites in the C14 hexagonal Laves phase [34], it was assumed that all hydrogen atoms occupy the sites with the Wyckoff symbol 12*k*, because these sites do not share the faces of the interstitial tetrahedra and thus the hydrogen atoms occupying them are not too close to each other [34]. These assumptions uniquely determine the hydrogen-occupied sites for each supercell model. Fig. 1b shows one of the thus obtained models considered in the present *ab initio* simulations.

Table 1. Numbers of tetrahedral interstitial sites with different local environments per formula unit and ratios of their individual volumes to alloy volume per formula unit for AB$_2$ Laves phases with ideal atomic positions.

| Tetrahedra Site | Number | Volume Ratio |
| --- | --- | --- |
| B$_4$ | 1 | 1/24 (~0.0417) |
| AB$_3$ | 4 | 5/96 (~0.0521) |
| A$_2$B$_2$ | 12 | 1/16 (= 0.0625) |

**2.2.3. Hydrogen binding energy calculations**

*Ab initio* density functional theory (DFT) calculations were performed using the VASP code [50-52] with the plane-wave basis projector augmented wave (PAW) method [53]. The exchange-correlation energy was obtained within the generalized gradient approximation (GGA) of the Perdew-Burke-Ernzerhof (PBE) form [54]. The plane-wave cutoff energy was set to 400 eV. Reciprocal spaces were sampled by a Γ-centered 4 × 4 × 6 *k*-point mesh for the 48-metal-atom supercell models and the Methfessel-Paxton method [55] with the smearing width of 0.1 eV. The 3d 4s orbitals of titanium, chromium, manganese, iron, and nickel and the 4s 4p 4d 5s orbitals of zirconium were treated as the valence states. Total energies were minimized until they converged within 1 × 10$^{-3}$ eV per simulation cell for each ionic step. All calculations were performed by considering spin polarization, a fact that was also experimentally examined by magnetic measurements, as discussed in Appendix A and Fig. A.

To obtain the energy-volume curves, seven volumes in the ranges of 144 to 180 Å$^3$/u.c. and 186 to 222 Å$^3$/u.c. (u.c.: unit cell) were considered for the systems with and without hydrogen atoms, respectively. For each composition, the obtained energies of six SQS-based models for the given volumes were then fitted to the Vinet equation of state [56,57] to obtain the volume and the energy in the equilibrium state. Metal and hydrogen atoms were initially put on the ideal Laves-phase lattice sites and the geometric centers (centroids) of the A$_2$B$_2$ 12*k* interstitial sites, respectively. The atomic positions were then relaxed with fixing the cell shape and volume until all the forces on the atoms converged within 5 × 10$^{-2}$ eV/Å. The hydrogen binding energy per atom was then computed as

$$\Delta E_\text{H} = \frac{1}{3}\left[E(\text{AB}_2\text{H}_3) - E(\text{AB}_2) - \frac{3}{2}E(\text{H}_2)\right] \quad (1)$$

where $E(\text{H}_2)$, $E(\text{AB}_2)$, and $E(\text{AB}_2\text{H}_3)$ are the energies of H$_2$, AB$_2$, and AB$_2$H$_3$ per formula unit, respectively. The energy of the H$_2$ molecule was computed in a 20 × 20 × 20 Å$^3$ simulation cell and for the Γ point in the reciprocal space. The obtained hydrogen-hydrogen distance was 0.751 Å, in reasonable agreement with the experimental value of 0.74144 Å [58].

**2.3. Experimental Procedures**

The ingots of HEAs with compositions $Ti_xZr_{2-x}CrMnFeNi$ ($x$ = 0.4, 0.6, 1.2 and 1.6) with ~10 g mass were prepared by arc melting using pieces of pure titanium (99.99%), zirconium (99.5%), chromium (99.99%), manganese (99.95%), iron (99.97%) and nickel (99.9%). The pieces were melted and mixed in a water-cooled copper crucible under an argon atmosphere. To increase the homogeneity of the alloys, the mixture was remelted six times. Ingots produced by arc melting were characterized by various methods, as described below.

First, the crystal structure was examined by X-ray diffraction (XRD) using Cu Kα irradiation with a filament current of 40 mA and an acceleration voltage of 45 kV. The Rietveld method using the PDXL software was used to identify the phases and determine their lattice parameters.

Second, the microstructure of the samples was investigated by a scanning electron microscope (SEM) equipped with energy-dispersive X-ray spectroscopy (EDS) and electron backscatter diffraction (EBSD) at 15 kV. The samples for SEM were prepared by cutting a piece of the ingot using electric discharge machining, followed by mechanical grinding using sandpapers, polishing using buff and 9 µm and 3 µm diamond suspensions, and final polishing by buff and colloidal silica with 60 nm particle size.

Third, the nanostructure of the samples was examined using transmission electron microscopy (TEM) and scanning-transmission electron microscopy (STEM) at 200 kV using high-resolution imaging, fast Fourier transform (FFT) analysis and EDS analysis. For TEM and STEM analyses, a small piece of the samples was crushed in ethanol and dispersed on a carbon grid.

Fourth, hydrogen storage was examined using a Sieverts-type gas absorption apparatus at room temperature (298 K). For this experiment, ~1 g of the alloys was crushed in an air atmosphere to achieve particle sizes below 75 µm. The crushed samples were examined in terms of pressure-composition-temperature (PCT) isotherms for three hydriding-dehydriding cycles followed by a hydrogenation kinetic measurement for one cycle. The samples before PCT measurements were subjected to evacuation at room temperature for 3 h to remove air and moisture, but no thermal activation treatment was performed. The crystal structure of the hydrides after hydrogen storage was also examined by conducting the XRD analysis immediately after hydriding. Moreover, cyclic hydrogenation-dehydrogenation measurements were conducted for 1000 cycles (10 min hydrogenation under an initial hydrogen pressure of 3.7 MPa followed by 20 min evacuation at 298 K) to examine the cycling stability of the alloys.

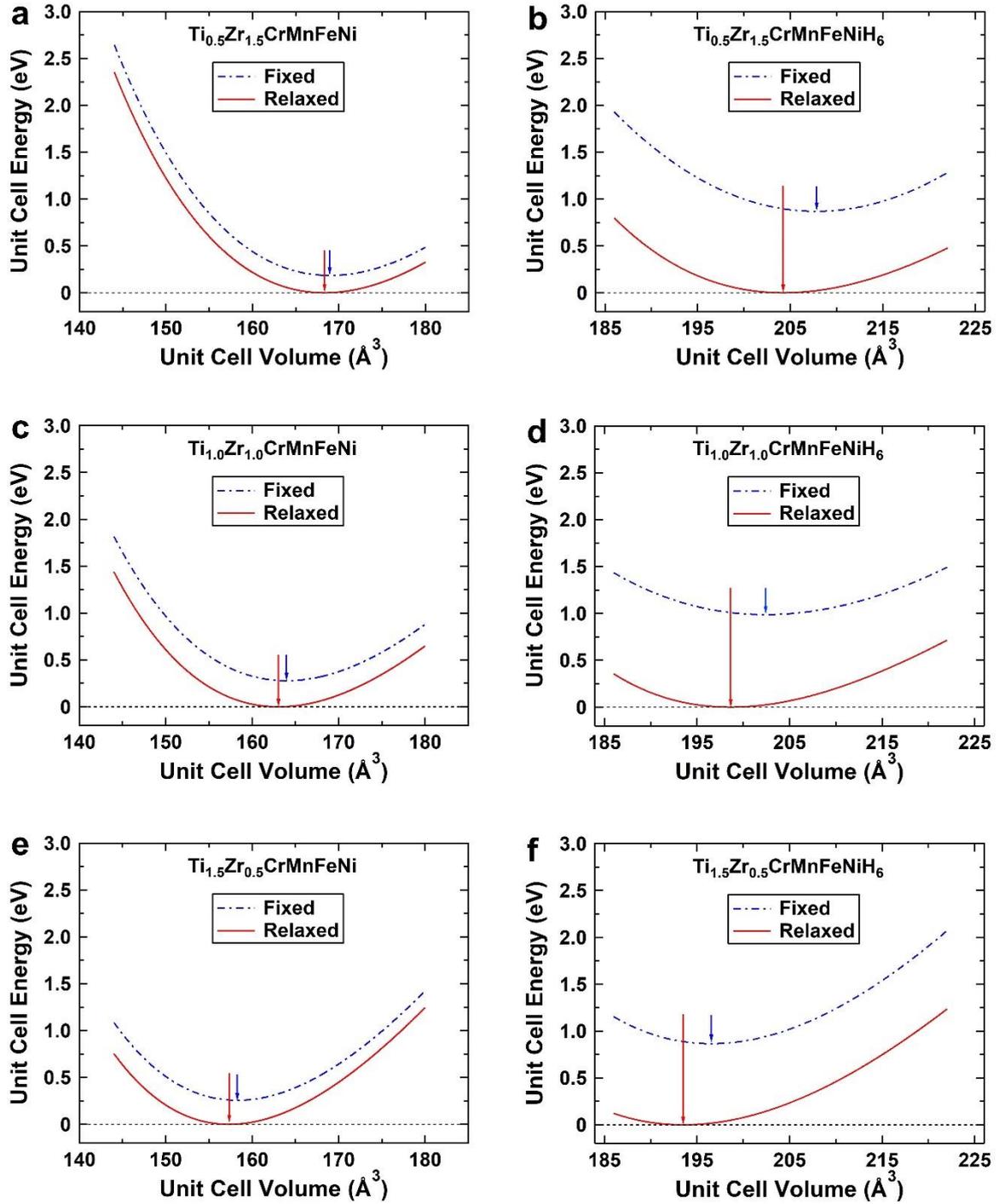

Fig. 2. *Ab initio* calculated energy-volume curves fitted to the Vinet equation of state without and with relaxation of atomic positions for (a, c, e) hydrogen-free HEAs Ti$_x$Zr$_{2-x}$CrMnFeNi and (b, d, f) corresponding high-entropy hydrides. Arrows show equilibrium volumes for fixed and relaxed atomic positions.

## 3. Results
### 3.1. Calculated hydrogen binding energy

Figs. 2a, c and e show the *ab initio* computed energy-volume curves before and after relaxation of atomic positions for the Ti$_x$Zr$_{2-x}$CrMnFeNi alloys with $x = 0.5$, 1.0 and 1.5, respectively.

The equilibrium unit cell volume decreases with increasing the fraction of titanium. Moreover, for all the compositions, the equilibrium unit cell volume decreases after the relaxation, as summarized in Table 2. Fig. 2b, d and f show the *ab initio* computed energy-volume curves before and after relaxation of atomic positions for the Ti$_x$Zr$_{2-x}$CrMnFeNiH$_6$ hydrides with $x$ = 0.5, 1.0 and 1.5, respectively. In these hydrides, hydrogen atoms occupy all the 12$k$ A$_2$B$_2$ interstitial sites. Similar to the HEAs, the unit cell volume of the hydride decreases with increasing the titanium fraction, but as summarized in Table 2, the main difference is that the unit cell volumes for the hydrides are 21-23% larger compared to the corresponding alloys (i.e., larger by 3 Å$^3$ per hydrogen atom). Such a volume expansion is in a similar order to those experimentally reported for various binary and ternary hydrides [59] and also similar to the one previously calculated by *ab initio* for the Zr(Cr$_{0.5}$Ni$_{0.5}$)$_2$ alloy with the C14 Laves phase (3.2 Å$^3$ per hydrogen atom) [38]. Another difference between the alloys and hydrides is that the impact of atomic-position relaxation on the equilibrium volume and energy changes is much larger for the hydrides compared to the alloys. This is likely because the positions of metal atoms more strongly deviate from the ideal lattice sites due to the presence of hydrogen atoms, as detailed in Appendix B and Fig. B.

The hydrogen binding energy per hydrogen atom for relaxed atomic positions is given in Table 2 for the three high-entropy hydrides Ti$_x$Zr$_{2-x}$CrMnFeNiH$_6$. The hydrogen binding energies are -0.126, -0.105 and -0.074 eV per hydrogen atom for $x$ = 0.5, 1.0 and 1.5, respectively. These values are substantially lower than those at the A$_2$B$_2$ sites in the Laves phases consisting of similar elements but chemically less disordered such as -0.27 eV per hydrogen atom for TiCr$_2$ [36], and -0.20 to -0.28 eV per hydrogen atom for Zr(Cr$_{0.5}$Ni$_{0.5}$)$_2$ [38]. This implies that the present Laves HEAs can more easily desorb the hydrogen atoms, and thus, can be more practical for applications where low-temperature hydrogen storage is needed. Fig. 3 visualizes the dependence of the hydrogen binding energy as a function of the titanium content for relaxed atomic positions. With increasing the titanium content, the hydrogen binding energy becomes less negative, and thus the hydrides become less stable. Therefore, it is expected that the alloys with lower titanium content can desorb hydrogen at lower temperatures and lower pressures. The hydrogen binding energies for Ti$_{0.5}$Zr$_{1.5}$CrMnFeNiH$_6$ and Ti$_{1.0}$Zr$_{1.0}$CrMnFeNiH$_6$ are slightly more negative than -0.1 eV, suggesting the potential of these two alloys for room-temperature hydrogen storage [29]. The strong dependence of the hydrogen binding energy on the composition also indicates the possibility to tailor the properties of hydrogen storage alloys by modifying the compositions of HEAs [60].

Table 2. *Ab initio* calculated equilibrium volumes for fixed and relaxed atomic positions, unit cell energy difference between fixed and relaxed conditions, and hydrogen binding energy for HEAs Ti$_x$Zr$_{2-x}$CrMnFeNi and their corresponding Ti$_x$Zr$_{2-x}$CrMnFeNiH$_6$ hydrides where hydrogen atoms occupy all the 12$k$ A$_2$B$_2$ interstitial sites.

| Alloy/Hydride | Volume (Å$^3$/u.c.) | | $E_{Relaxed}$-$E_{Fixed}$ (Å$^3$/u.c.) | $\Delta E_H$ (eV / H atom) |
|---|---|---|---|---|
| | Fixed | Relaxed | | |
| Ti$_{0.5}$Zr$_{1.5}$CrMnFeNi | 169.0 | 168.3 | -0.18 | |
| Ti$_{1.0}$Zr$_{1.0}$CrMnFeNi | 164.0 | 163.0 | -0.28 | |
| Ti$_{1.5}$Zr$_{0.5}$CrMnFeNi | 158.3 | 157.3 | -0.26 | |
| Ti$_{0.5}$Zr$_{1.5}$CrMnFeNiH$_6$ | 207.8 | 204.2 | -0.87 | - 0.126 |
| Ti$_{1.0}$Zr$_{1.0}$CrMnFeNiH$_6$ | 202.4 | 198.6 | -0.99 | - 0.105 |
| Ti$_{1.5}$Zr$_{0.5}$CrMnFeNiH$_6$ | 196.5 | 193.5 | -0.87 | - 0.074 |

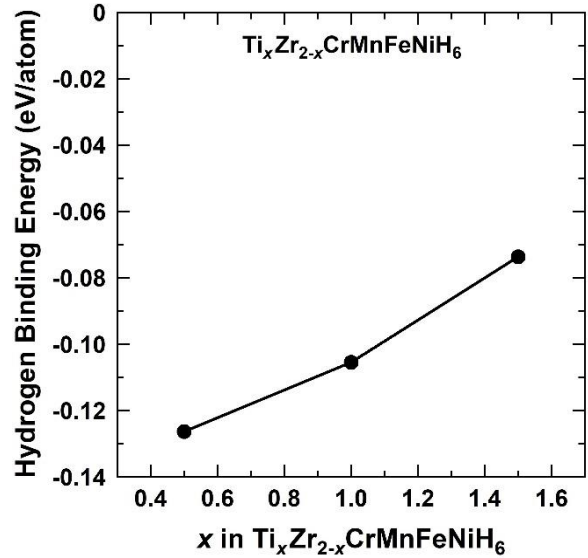

Fig. 3. *Ab initio* calculated hydrogen binding energy as a function of titanium content for high-entropy hydrides $Ti_xZr_{2-x}CrMnFeNiH_6$. Atomic positions were relaxed for these calculations.

### 3.2. Structural features of alloys and hydrides

Fig. 4 summarizes the results of the XRD analysis. Fig. 4a shows the XRD profiles of the HEAs, $Ti_xZr_{2-x}CrMnFeNi$ ($x$ = 0.4-1.6). All samples have mainly a hexagonal C14 Laves phase structure with the $P6_3/mmc$ space group, while the presence of weak peaks at 42-43° for the Ti-rich alloys suggests that minor amounts of a cubic phase can also be present [27]. Fig. 4b shows a magnified view of the XRD profiles for the (110) peak of the Laves phase located between 35.5° and 37.5 °. This figure demonstrates a systematic shift of this peak to higher angles with increasing the titanium fraction, indicating that a lattice contraction occurs by increasing the titanium fraction. The occurrence of lattice contraction by increasing the titanium fraction is shown more clearly in Table 3, where the calculated lattice parameters using the Rietveld method are given. It should be noted that the plots of calculated-observed intensities in the Rietveld analysis indicate a maximum intensity difference of 10%, which is a reasonable level. Such a difference should be partly due to the inherent distortion in high-entropy materials [7] and partly due to the presence of a small amount of a cubic phase [27]. Fig. 4c compares the XRD pattern of $Ti_{0.4}Zr_{1.6}CrMnFeNi$ before and after hydriding. The hydride phase also has a hexagonal C14 crystal structure, while the shift of the XRD profile to lower angles after hydriding suggests that the lattice is expanded in the presence of hydrogen atoms in the lattice. Fig. 4d summarizes the lattice parameters calculated from the XRD analysis for both alloys and hydrides. Note that the data for $Ti_{1.0}Zr_{1.0}CrMnFeNi$ was calculated from the XRD profiles reported in an earlier publication [27]. It should be also noted that the crystal structure of the $Ti_{1.2}Zr_{0.8}CrMnFeNiH_6$ and $Ti_{1.6}Zr_{0.4}CrMnFeNiH_6$ hydrides could not be identified because these two hydrides released hydrogen very fast under ambient conditions and before the XRD analysis. Both *a* and *c* lattice parameters increase with increasing the number of zirconium atoms, which have a larger atomic radius than titanium atoms [61]. Moreover, the lattice parameters increase with the addition of hydrogen atoms to the materials.

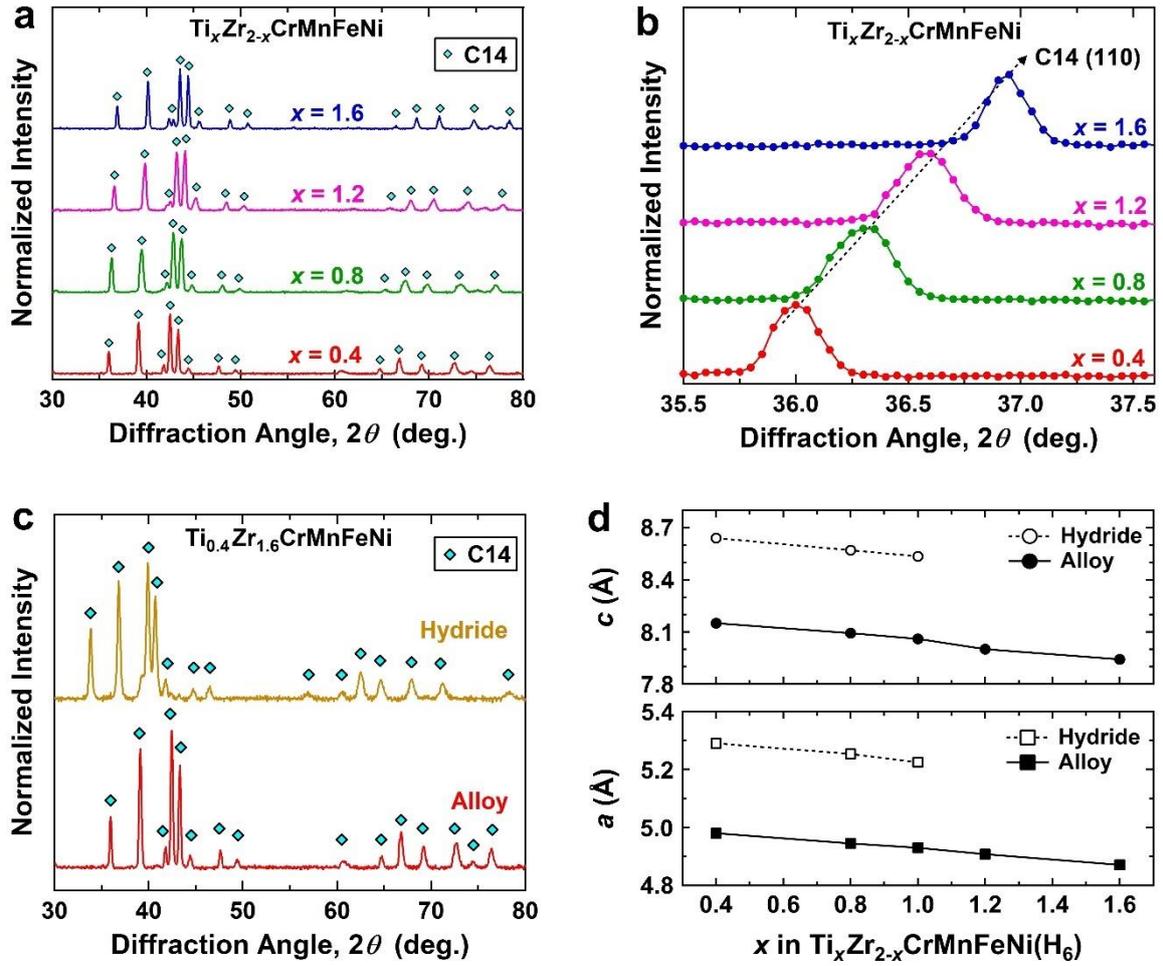

Fig. 4. (a, b) XRD profiles of hydrogen-free HEAs Ti$_x$Zr$_{2-x}$CrMnFeNi where (b) shows magnified view of (110) peak of C14 Laves phase, (c) XRD profiles of hydrogen-free HEA Ti$_{0.4}$Zr$_{1.6}$CrMnFeNi and corresponding high-entropy hydride, and (d) lattice parameters of hydrogen-free HEAs Ti$_x$Zr$_{2-x}$CrMnFeNi and corresponding high-entropy hydrides determined from XRD profiles using Rietveld method. Data for Ti$_{1.0}$Zr$_{1.0}$CrMnFeNi in (d) were taken from literature [27].

Table 3. Lattice parameters of C14 Laves phase obtained experimentally using XRD analysis for HEAs Ti$_x$Zr$_{2-x}$CrMnFeNi.

| Alloy | $a$ (Å) | $c$ (Å) |
|---|---|---|
| Ti$_{0.4}$Zr$_{1.6}$CrMnFeNi | 4.982 ± 0.008 | 8.148 ± 0.020 |
| Ti$_{0.8}$Zr$_{1.2}$CrMnFeNi | 4.945 ± 0.014 | 8.093 ± 0.030 |
| Ti$_{1.2}$Zr$_{0.8}$CrMnFeNi | 4.908 ± 0.010 | 8.002 ± 0.020 |
| Ti$_{1.6}$Zr$_{0.4}$CrMnFeNi | 4.871 ± 0.011 | 7.942 ± 0.020 |

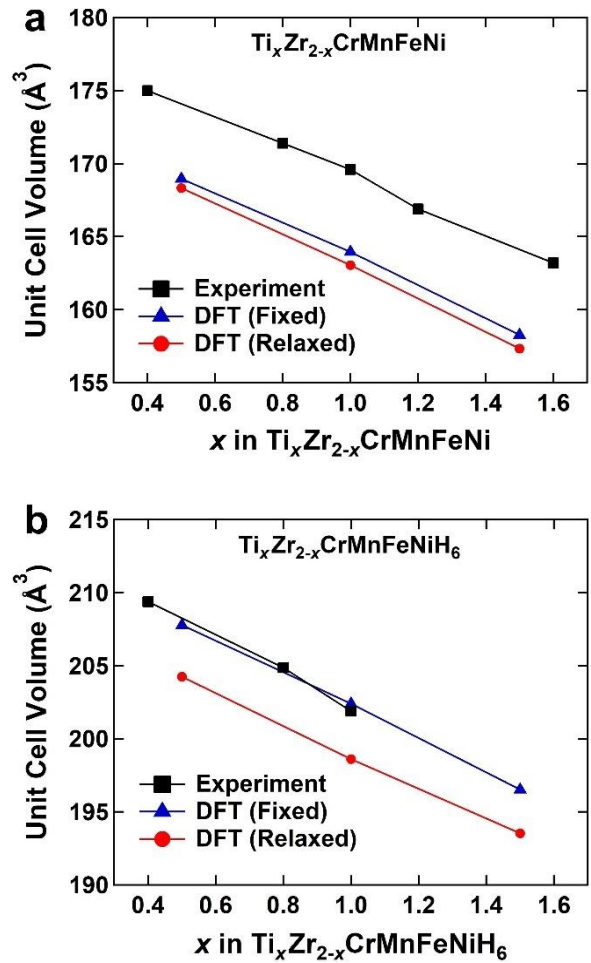

Fig. 5. Unit cell volume of C14 Laves phase obtained experimentally using XRD analysis and calculated using *ab initio* without (fixed) and with relaxation of atomic positions for (a) HEAs Ti$_x$Zr$_{2-x}$CrMnFeNi and (b) corresponding high-entropy hydrides.

It is now possible to compare the unit cell volumes obtained experimentally and theoretically to validate the models used for the first-principles calculations. Fig. 5 compares the equilibrium volumes obtained by DFT with those obtained experimentally for the (a) alloys and (b) hydrides with different fractions of titanium. The experimental and theoretical data show similar trends and the unit cell volume decreases with increasing the titanium content for both alloys and hydrides. The linear change in the unit cell volume is qualitatively consistent with Vegard's law by considering the difference in the atomic radius of titanium and zirconium [61]. The *ab initio* calculated volumes are slightly smaller than those obtained in experiments: 3.5% for the alloys and 2.5% for the hydrides. Such an underestimation of volume using GGA (like PW91 [62,63] or PBE [54]) is found often in systems containing magnetic 3d elements such as pure bcc iron [64-66] and the HEA CrMnFeCoNi [67,68]. The small differences between the experimental and theoretical cell volumes confirm the validity of the models used for the simulation of the HEAs and hydrides.

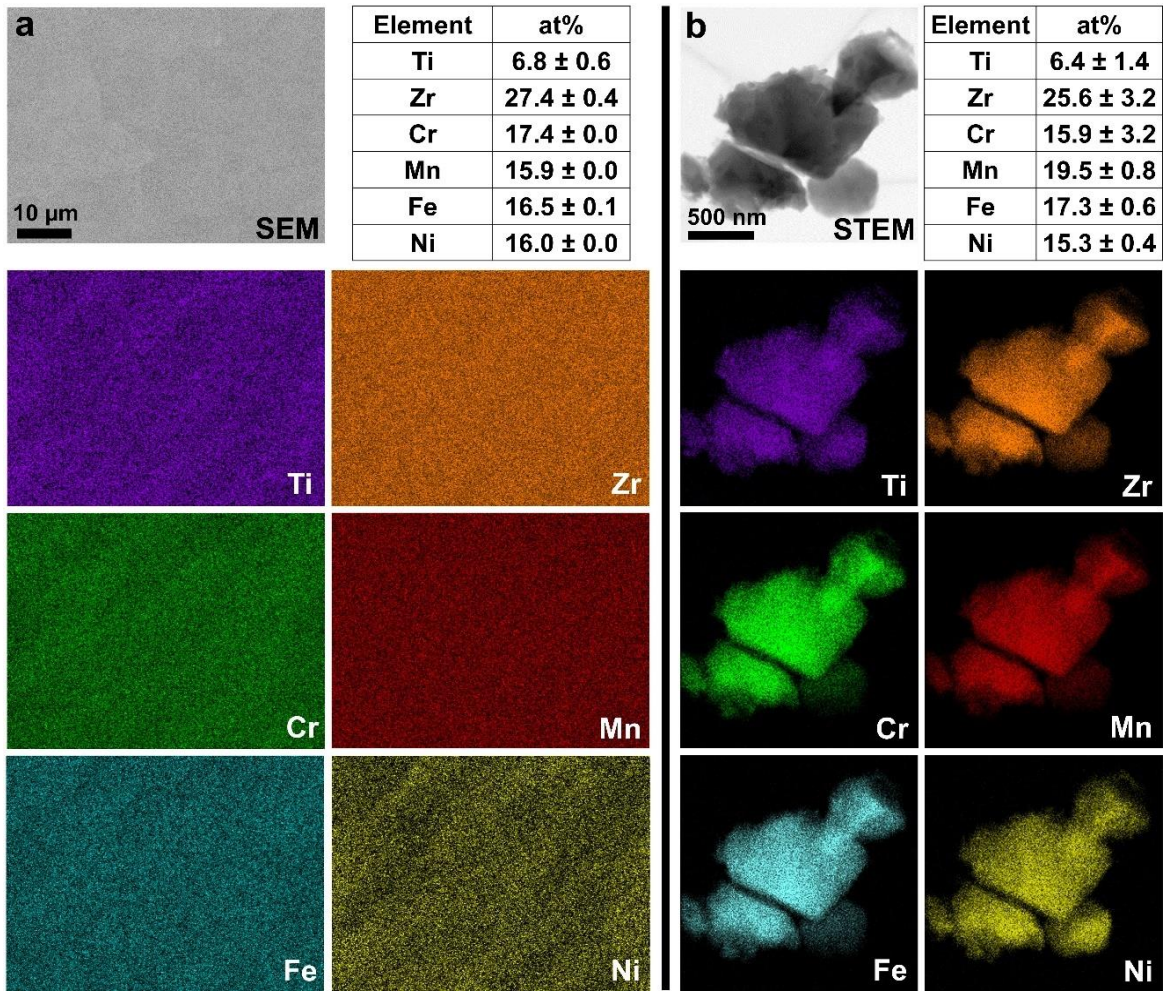

Fig. 6. Distribution of elements and their atomic fractions in HEA $Ti_{0.4}Zr_{1.6}CrMnFeNi$ examined by EDS elemental mappings using (a) SEM and (b) STEM.

### 3.3. Microstructure of alloys

The microstructure of $Ti_{0.4}Zr_{1.6}CrMnFeNi$, taken as a representative alloy with the lowest hydrogen binding energy, is shown in Figs. 6 and 7. Fig. 6 shows (a) SEM-EDS and (b) STEM-EDS analyses. The distribution of elements is reasonably uniform at the micrometer and nanometer levels. Moreover, the fraction of elements is reasonably consistent with the nominal composition within the detection limits of EDS analysis. The EDS analysis confirms that arc melting can be successfully employed to synthesize the HEAs. An SEM micrograph and corresponding EBSD crystal orientation and phase mappings are shown in Figs. 7a-c, respectively. Fig. 7b illustrates that the HEA includes mainly large and elongated grains with sizes of several hundred micrometers and Fig. 7c confirms that the alloy contains mainly a C14 Laves phase in good agreement with the XRD analysis. Figs. 7d-f show the high-resolution TEM images of the HEA. The C14 Laves phase was the only phase that could be detected by high-resolution TEM images in good agreement with the XRD and EBSD analyses. Most of the examined regions such as the one in Fig. 7d are free from defects, while high-angle grain boundaries and dislocations are visible in some regions, as shown in Figs. 7e and 7f, respectively. These microstructural features are rather similar to the microstructures of other Laves phases synthesized by arc melting [15,16,27].

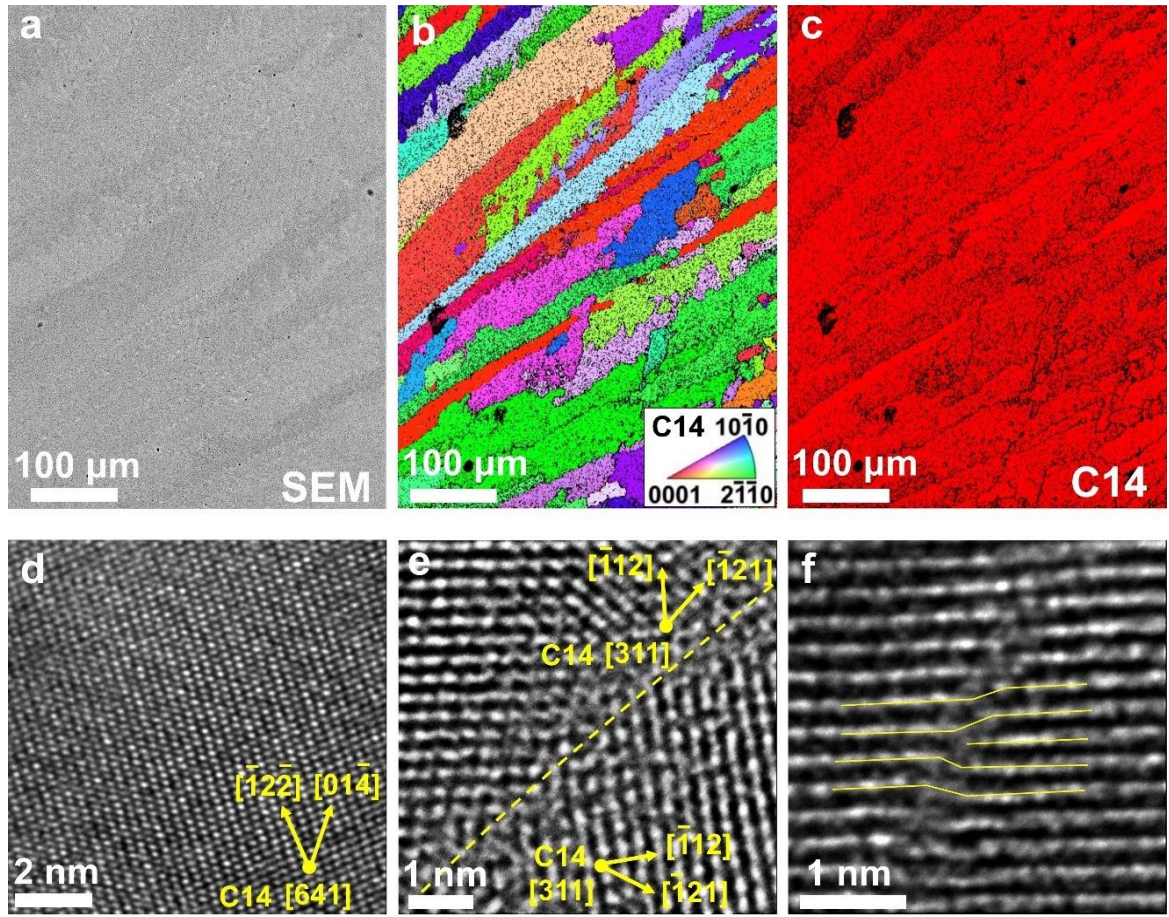

Fig. 7. (a) SEM micrograph and corresponding (b) crystal orientation and (c) phase mappings achieved by EBSD with beam step size of 1 μm; and (d-f) high-resolution TEM lattice images for (d) C14 Laves phase lattice, (e) high-angle grain boundary and (f) dislocation for HEA $Ti_{0.4}Zr_{1.6}CrMnFeNi$.

### 3.4. Room temperature hydrogen storage

Figs. 8a-d show the PCT absorption/desorption isotherms for three cycles at 298 K for the $Ti_xZr_{2-x}CrMnFeNi$ alloys with $x = 0.4, 0.8, 1.2$ and $1.6$, respectively. Fig. 8a indicates that the $Ti_{0.4}Zr_{1.6}CrMnFeNi$ alloy reversibly absorbs and desorbs 1.7 wt% of hydrogen with good cycling performance. The maximum hydrogen-to-metal ($H/M$) ratio for the alloy reaches 1.06, confirming that the composition of the hydride can be reasonably considered as $Ti_{0.4}Zr_{1.6}CrMnFeNiH_6$. PCT isotherms for the $Ti_{0.8}Zr_{1.2}CrMnFeNi$ alloy in Fig. 8b show a similar trend as for $Ti_{0.4}Zr_{1.6}CrMnFeNi$. Further, the alloy can store 1.6 wt% of hydrogen with a hydrogen-to-metal ratio of 1, corresponding to the $Ti_{0.8}Zr_{1.2}CrMnFeNiH_6$ hydride. The main difference between Figs. 8a and 8b is that the alloy with less titanium exhibits a lower plateau pressure which should be due to its stronger hydrogen binding energy, as demonstrated in Fig. 3 using *ab initio* simulations. Fig. 8c shows the PCT isotherms for the $Ti_{1.2}Zr_{0.8}CrMnFeNi$ sample. This HEA absorbs 1.4 wt% hydrogen with a hydrogen-to-metal ratio of 0.8. Since the plateau pressure for this alloy is high, its complete hydrogenation does not occur by increasing the pressure to 9 MPa (i.e., the upper limit of pressure in the authors' gas absorption facility). As shown in Fig. 8d, the $Ti_{1.6}Zr_{0.4}CrMnFeNi$ alloy absorbs a minor amount of 0.1 wt% hydrogen, suggesting that the plateau pressure of this alloy is higher than 9 MPa because of its weak hydrogen binding energy, as discussed in Fig. 3.

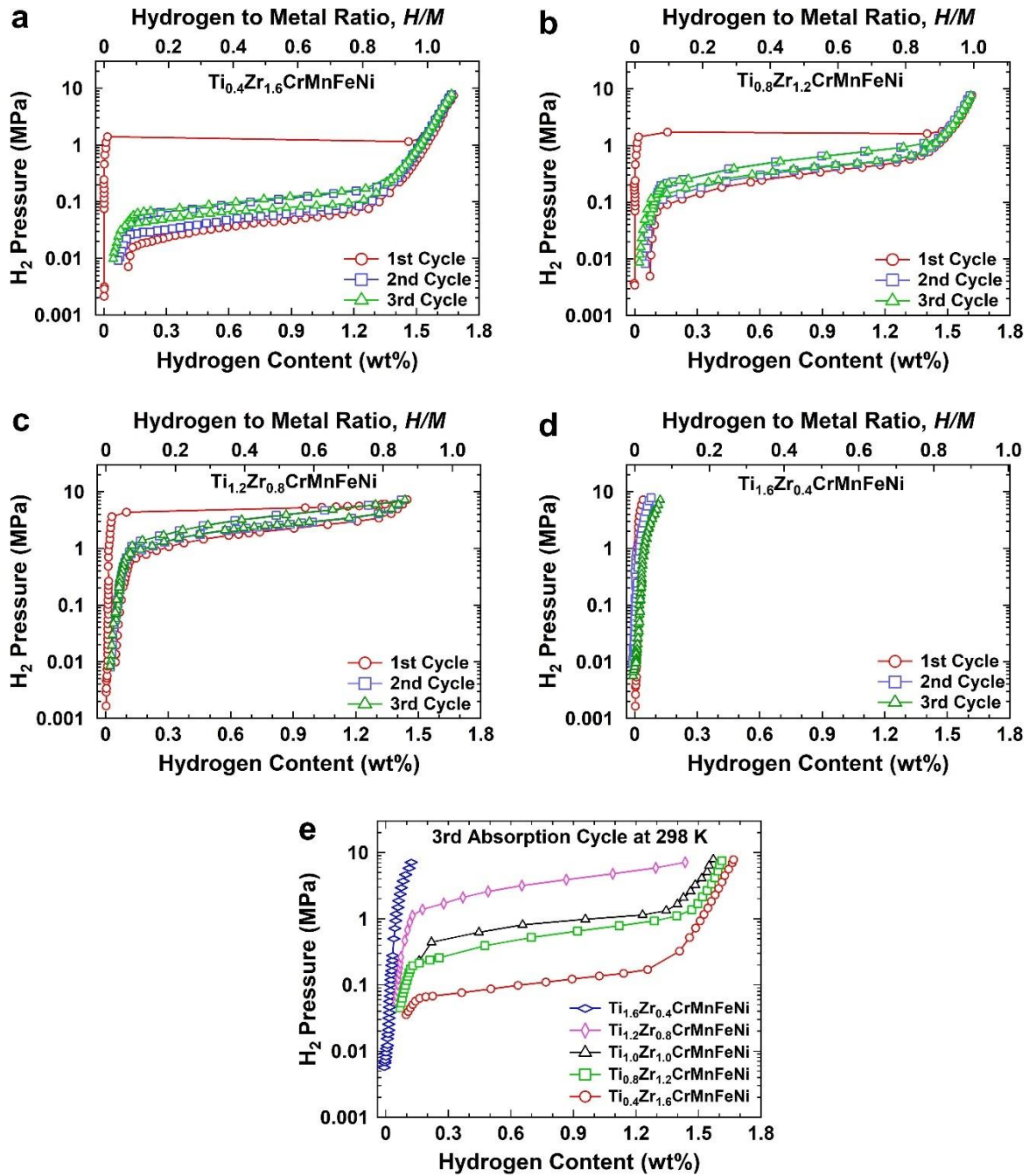

Fig. 8. PCT absorption/desorption isotherms at room temperature for HEAs (a) $Ti_{0.4}Zr_{1.6}CrMnFeNi$, (b) $Ti_{0.8}Zr_{1.2}CrMnFeNi$, (c) $Ti_{1.2}Zr_{0.8}CrMnFeNi$ and (d) $Ti_{1.6}Zr_{0.4}CrMnFeNi$; and (e) comparison of third cycle of PCT absorption isotherms for HEAs. Data for $Ti_{1.0}Zr_{1.0}CrMnFeNi$ in (e) were taken from literature [27].

Fig. 8e provides a clear comparison between the PCT isotherms in the third absorption cycle for the $Ti_xZr_{2-x}CrMnFeNi$ alloys with $x$ = 0.4, 0.8, 1.0, 1.2 and 1.6, where the data for the Ti$_{1.0}$Zr$_{1.0}$CrMnFeNi alloy were taken from the literature [27]. The plateau pressure systematically increases by increasing the fraction of titanium, while the Ti$_{0.4}$Zr$_{1.6}$CrMnFeNi and Ti$_{0.8}$Zr$_{1.2}$CrMnFeNi alloys have plateau pressures close to ambient pressure which renders them appropriate for practical applications. Such an increase in the plateau pressure by increasing the titanium content can be well explained by the influence of titanium on the hydrogen binding energy, as shown in Fig. 3.

Fig. 9 demonstrates the kinetic measurements of hydrogen storage under an initial hydrogen pressure of 3.7 MPa at room temperature for the HEAs. The alloys absorb hydrogen very fast within almost 30 seconds, although the total amount of hydrogen decreases with increasing the fraction of titanium, i.e., with increasing the plateau pressure. The amount of stored hydrogen in Ti$_{0.4}$Zr$_{1.6}$CrMnFeNi in both kinetic and PCT measurements is higher than the capacity of commercial room-temperature hydrogen storage materials [1-6].

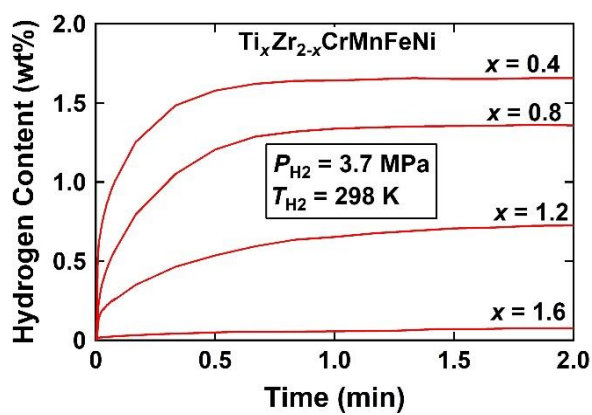

Fig. 9. Hydrogenation kinetic curves at room temperature for HEAs Ti$_x$Zr$_{2-x}$CrMnFeNi.

Fig. 10 summarizes the cyclic hydrogen storage measurements for the Ti$_{0.4}$Zr$_{1.6}$CrMnFeNi for up to 1000 cycles, where (a) shows the amount of hydrogen absorbed in each cycle within 10 min under an initial hydrogen pressure of 3.7 MPa at 298 K and (b) shows the corresponding PCT absorption/desorption isotherms for cycles 4, 30, 100 and 1000. The amount of absorbed hydrogen remains almost constant within 1000 cycles, and the PCT isotherm also does not show any significant change after 1000 hydrogenation-dehydrogenation cycles. The current results confirm the excellent cycling performance and stability of this HEA, which is a critical issue in the commercialization of hydrogen storage materials [1].

## 4. Discussion

HEAs are the most recent alloys that have been explored for hydrogen storage [8-25]. The presence of multi-principal elements in the lattice of these alloys makes it possible to tune their electronic structures, crystal structures and physical properties in a much more straightforward way as compared to conventional alloys and intermetallics [7]. Since the capability to store hydrogen strongly depends on the electronic structure [1-3], the flexibility in controlling the electronic structure of HEAs introduces them as potential materials for hydrogen storage applications. However, the investigations on high-entropy hydrogen storage materials are still in their infancy, and only a few attempts have been pursued to combine theoretical calculations and experiments to develop new

HEAs with appropriate electronic structures for room-temperature hydrogen storage [7]. The current study addresses this challenge and combines the existing empirical knowledge on the development of HEAs for low-temperature hydrogen storage with first-principles calculations to design alloys that can reversibly store hydrogen at ambient temperature under pressures close to atmospheric pressure. The empirical knowledge suggests that HEAs with $AB_2$-type Laves phase structure and a VEC close to 6.4 have a high potential for low-temperature hydrogen storage [8,15,16,27]. Previous first-principles calculations on conventional Mg-based alloys also suggest that a hydrogen binding energy of -0.1 eV or slightly more negative can lead to room-temperature hydrogen storage [26]. The $Ti_xZr_{1-x}CrMnFeNi$ alloys were therefore selected for the present study because they satisfy all empirical requirements and because their theoretical hydrogen binding energy can be tuned to negative values smaller than -0.1 eV by changing the fraction of titanium and zirconium atoms.

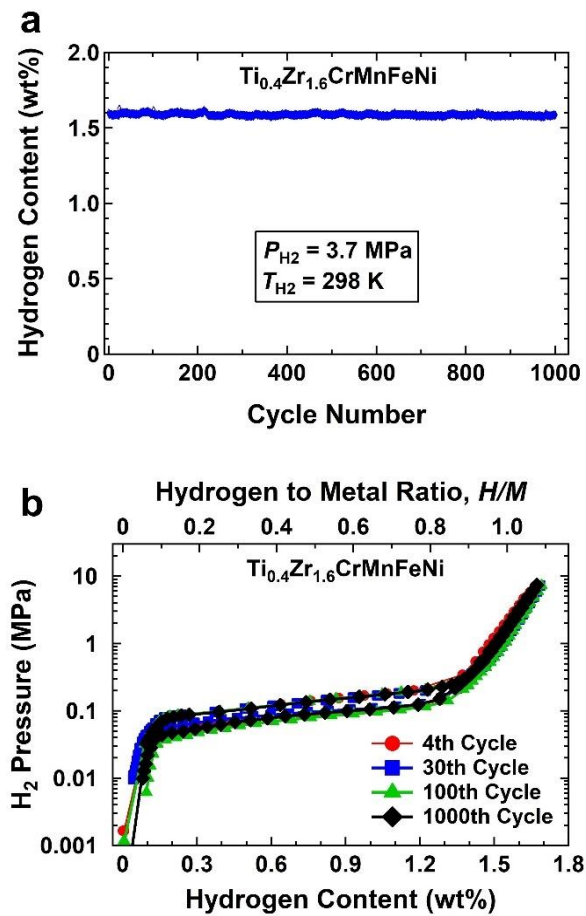

Fig. 10. (a) Hydrogenation cycling tests for 1000 cycles at room temperature and (b) corresponding PCT absorption/desorption isotherms for cycles 4, 30, 100 and 1000 for HEA $Ti_{0.4}Zr_{1.6}CrMnFeNi$.

The designed alloys reversibly adsorb and desorb hydrogen at room temperature, while the equilibrium pressure can be easily tuned to appropriately low values by reducing the hydrogen binding energy to more negative values via increasing the fraction of zirconium. The kinetics of hydrogen storage is quite fast in these alloys due to their Laves phase structure [29]. Moreover, such fast storage kinetics occur without any thermal activation or catalyst addition, in contrast to many other storage materials including TiFe which suffer from the activation problem at room temperature [3,5]. Another advantage is that the good hydrogen storage performance of the present HEAs was achieved even though they were handled and crushed under an air atmosphere. Such a good air resistance cannot be achieved easily for other room-temperature hydrogen storage materials without conducting chemical modification [3] mechanochemical process [69] or mechanical treatment through severe plastic deformation [70]. Among the selected alloys in this study, the amount of stored hydrogen in $Ti_{0.4}Zr_{1.6}CrMnFeNi$ and $Ti_{0.8}Zr_{1.2}CrMnFeNi$ alloys is higher than the capacity of commercial rare-earth-based $LaNi_5$ for room-temperature hydrogen storage [4,71,72]. The working pressure of these two alloys is also close to atmospheric pressure, in good agreement with their hydrogen binding energies, suggesting their high potential for stationary applications [1,2]. Therefore, a combination of first-principles calculations and experimental studies is an effective approach to design and synthesize new high-entropy hydrides for room-temperature hydrogen storage. Such an approach can also contribute to the advancement of nickel-metal-hydride (Ni-MH) batteries because HEAs were recently reported to have high potential as anode materials of Ni-MH batteries [73].

## 5. Conclusion

This study demonstrates the successful design and synthesis of high-entropy alloys having the capability to store hydrogen at room temperature and under pressures close to atmospheric pressures. The $AB_2$-type Laves phase $Ti_xZr_{2-x}CrMnFeNi$ ($x$ = 0.4-1.6) alloys with a valence electron concentration of 6.4 and low hydrogen binding energies (negative values close to -0.1 eV) were designed theoretically by using first-principles calculations and fabricated experimentally by the conventional arc melting method. These alloys reversibly absorb and desorb up to 1.7 wt% of hydrogen at room temperature (298 K) with fast kinetics, while their (de)hydrogenation pressure is systematically reduced by strengthening the hydrogen binding energy through increasing the zirconium fraction. To the authors' knowledge, this study is the first demonstration of adjusting the hydrogen storage temperature and pressure of high-entropy alloys to ambient conditions by employing the concept of binding-energy engineering. The concept introduced in the current study can be universally employed to discover many hydrogen storage materials for practical applications.

**Acknowledgments**

The authors thank Dr. Fritz Körmann of Max-Planck-Institut für Eisenforschung GmbH, Germany, and Prof. Ricardo Floriano of the University of Campinas, Brazil, for fruitful discussion. This work is supported in part by Grants-in-Aid for Scientific Research on Innovative Areas from the MEXT, Japan (JP19H05176 & JP21H00150), in part by the European Research Council (ERC) under the European Union's Horizon 2020 Research and Innovation Programme (grant agreement No 865855), in part by the State of Baden-Württemberg through bwHPC, and in part by the German Research Foundation (DFG) through grant number INST 40/467-1 FUGG (JUSTUS cluster).

## Appendix A

Fig. A shows the variations of magnetization versus magnetic field for (a) $Ti_{0.4}Zr_{1.6}CrMnFeNi$ and (b) $Ti_{1.6}Zr_{0.4}CrMnFeNi$ at 5 K and 300 K obtained using a superconducting quantum interference device (SQUID) magnetometer. Both alloys are paramagnetic at cryogenic and ambient temperatures and do not show a clear ferromagnetic behavior. The *ab initio* and experimental results may be comprehensively interpreted as follows; in the real alloys, each metal atom has a small magnetic moment on average, but they are randomly orientated even at cryogenic temperatures.

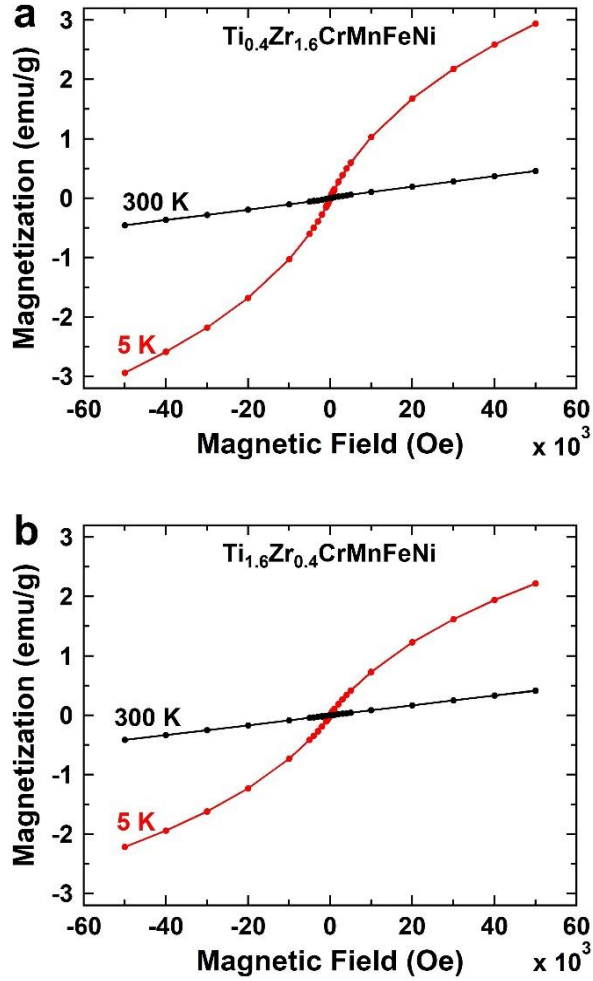

Fig. A. Magnetization versus magnetic field examined using SQUID at 5 K and 300 K for HEAs (a) $Ti_{0.4}Zr_{1.6}CrMnFeNi$ and (b) $Ti_{1.6}Zr_{0.4}CrMnFeNi$.

**Appendix B**

Atomic displacements are often considered to reflect the local lattice distortion in a HEA [74,75]. The variation of atomic displacements and lattice distortion in a HEA is particularly correlated with the magnitude of solid solution strengthening [76,77], although such distortions can also influence the functional properties of HEAs such as their activity for hydrogen storage [7-24]. Here, we quantify the atomic displacements from the ideal lattice sites for the $Ti_xZr_{2-x}CrMnFeNi$ alloys and $Ti_xZr_{2-x}CrMnFeNiH_6$ hydrides ($x$ = 0.5, 1.0 and 1.5) by *ab initio* DFT calculations. Fig. B shows the *ab initio* computed mean atomic displacements from the ideal lattice sites for each metal element as a function of unit cell volume. Overall, the atomic displacements become substantially larger after hydriding which is likely due to the presence of hydrogen atoms at interstitial sites. For the B-type elements, chromium shows the largest atomic displacements, followed by manganese, iron, and nickel. The magnitude of atomic displacements for the B-type elements is likely related to the binding energies to hydrogen atoms. This justification is consistent with the hydrogen binding energies in cubic C15 $ZrX_2$ alloys ($X$ = V, Cr, Mn, Fe, Co, Ni) reported in a previous *ab initio* study [35]. For the A-type elements, the atomic displacements of titanium increase with increasing the unit cell volume more strongly than for zirconium. This is likely because titanium has a smaller atomic radius than zirconium [61]. That is, at a large volume, titanium atoms have a larger space than zirconium atoms and thus can deviate from the ideal lattice sites more easily than zirconium atoms.

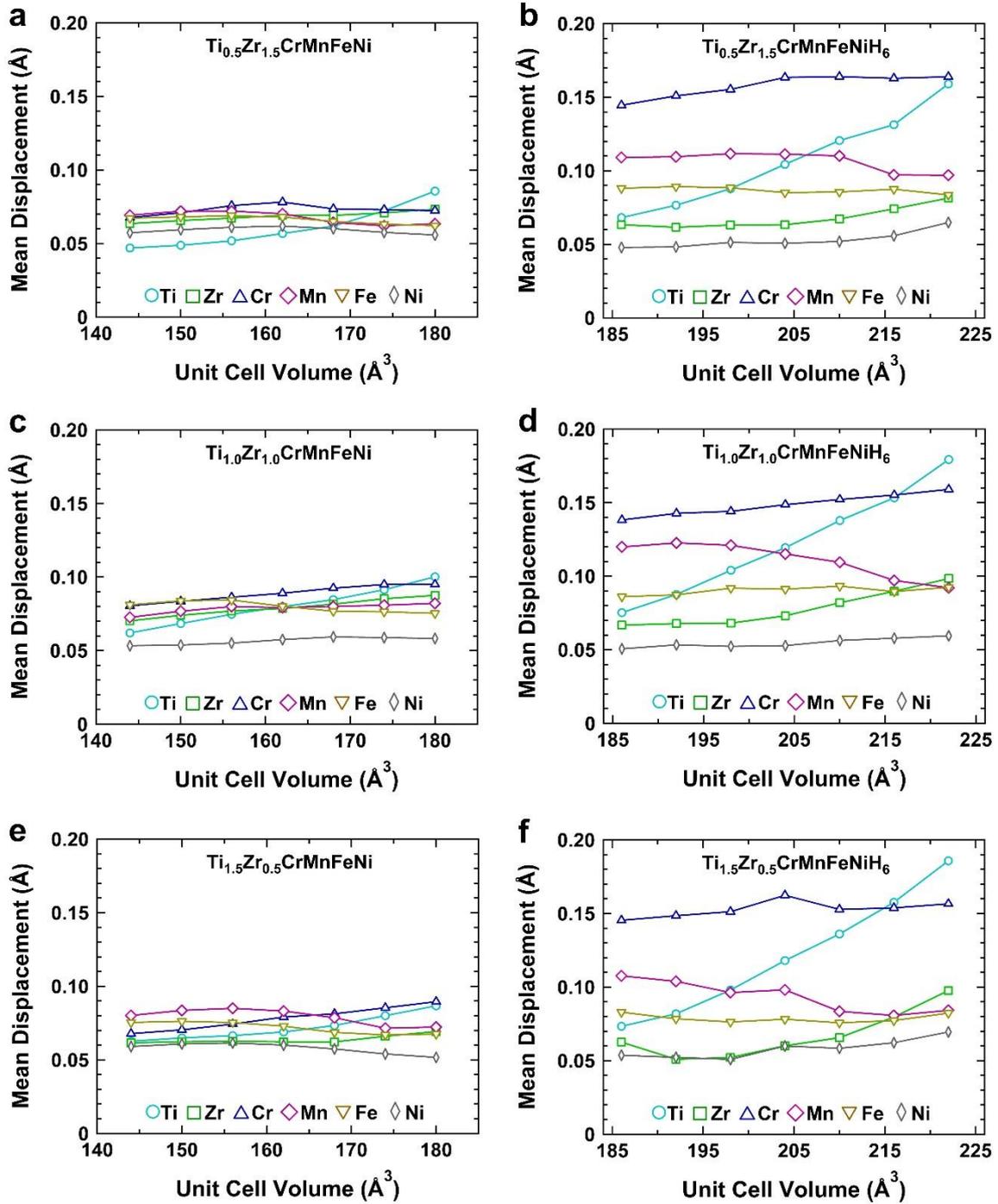

Fig. B. *Ab initio* calculated mean atomic displacements from ideal lattice sites for each metal element as a function of unit cell volumes after relaxation of atomic positions for (a, c, e) HEAs Ti$_x$Zr$_{2-x}$CrMnFeNi and (b, d, f) corresponding high-entropy hydrides.